\documentclass[]{aastex62}

\newcommand{\Msun}{\ensuremath{\mathrm{M}_\odot}}

\definecolor{pink}{RGB}{255, 20, 147}

\definecolor{carminered}{rgb}{1.0, 0.0, 0.22}

\definecolor{byzantine}{rgb}{0.74, 0.2, 0.64}

\definecolor{amber}{rgb}{1.0, 0.75, 0.0}

\definecolor{amethyst}{rgb}{0.6, 0.4, 0.8}

\definecolor{blue-violet}{rgb}{0.54, 0.17, 0.89}

\definecolor{comment}{RGB}{166, 38, 164}

\graphicspath{{./}{figures/}}

\submitjournal{ApJ Letters}

\usepackage{amsmath}
\usepackage{makecell}
\usepackage{multirow}
\usepackage{gensymb}
\begin{document}

\title{IceCube Search for Neutrinos Coincident with Compact Binary Mergers from LIGO-Virgo's First Gravitational-Wave Transient Catalog}

\collaboration{IceCube Collaboration}
\noaffiliation

\affiliation{III. Physikalisches Institut, RWTH Aachen University, D-52056 Aachen, Germany}
\affiliation{Department of Physics, University of Adelaide, Adelaide, 5005, Australia}
\affiliation{Dept. of Physics and Astronomy, University of Alaska Anchorage, 3211 Providence Dr., Anchorage, AK 99508, USA}
\affiliation{Dept. of Physics, University of Texas at Arlington, 502 Yates St., Science Hall Rm 108, Box 19059, Arlington, TX 76019, USA}
\affiliation{CTSPS, Clark-Atlanta University, Atlanta, GA 30314, USA}
\affiliation{School of Physics and Center for Relativistic Astrophysics, Georgia Institute of Technology, Atlanta, GA 30332, USA}
\affiliation{Dept. of Physics, Southern University, Baton Rouge, LA 70813, USA}
\affiliation{Dept. of Physics, University of California, Berkeley, CA 94720, USA}
\affiliation{Lawrence Berkeley National Laboratory, Berkeley, CA 94720, USA}
\affiliation{Institut f{\"u}r Physik, Humboldt-Universit{\"a}t zu Berlin, D-12489 Berlin, Germany}
\affiliation{Fakult{\"a}t f{\"u}r Physik {\&} Astronomie, Ruhr-Universit{\"a}t Bochum, D-44780 Bochum, Germany}
\affiliation{Universit{\'e} Libre de Bruxelles, Science Faculty CP230, B-1050 Brussels, Belgium}
\affiliation{Vrije Universiteit Brussel (VUB), Dienst ELEM, B-1050 Brussels, Belgium}
\affiliation{Dept. of Physics, Massachusetts Institute of Technology, Cambridge, MA 02139, USA}
\affiliation{Dept. of Physics and Institute for Global Prominent Research, Chiba University, Chiba 263-8522, Japan}
\affiliation{Dept. of Physics and Astronomy, University of Canterbury, Private Bag 4800, Christchurch, New Zealand}
\affiliation{Dept. of Physics, University of Maryland, College Park, MD 20742, USA}
\affiliation{Dept. of Astronomy, Ohio State University, Columbus, OH 43210, USA}
\affiliation{Dept. of Physics and Center for Cosmology and Astro-Particle Physics, Ohio State University, Columbus, OH 43210, USA}
\affiliation{Niels Bohr Institute, University of Copenhagen, DK-2100 Copenhagen, Denmark}
\affiliation{Dept. of Physics, TU Dortmund University, D-44221 Dortmund, Germany}
\affiliation{Dept. of Physics and Astronomy, Michigan State University, East Lansing, MI 48824, USA}
\affiliation{Dept. of Physics, University of Alberta, Edmonton, Alberta, Canada T6G 2E1}
\affiliation{Erlangen Centre for Astroparticle Physics, Friedrich-Alexander-Universit{\"a}t Erlangen-N{\"u}rnberg, D-91058 Erlangen, Germany}
\affiliation{Physik-department, Technische Universit{\"a}t M{\"u}nchen, D-85748 Garching, Germany}
\affiliation{D{\'e}partement de physique nucl{\'e}aire et corpusculaire, Universit{\'e} de Gen{\`e}ve, CH-1211 Gen{\`e}ve, Switzerland}
\affiliation{Dept. of Physics and Astronomy, University of Gent, B-9000 Gent, Belgium}
\affiliation{Dept. of Physics and Astronomy, University of California, Irvine, CA 92697, USA}
\affiliation{Karlsruhe Institute of Technology, Institut f{\"u}r Kernphysik, D-76021 Karlsruhe, Germany}
\affiliation{Dept. of Physics and Astronomy, University of Kansas, Lawrence, KS 66045, USA}
\affiliation{SNOLAB, 1039 Regional Road 24, Creighton Mine 9, Lively, ON, Canada P3Y 1N2}
\affiliation{Department of Physics and Astronomy, UCLA, Los Angeles, CA 90095, USA}
\affiliation{Department of Physics, Mercer University, Macon, GA 31207-0001, USA}
\affiliation{Dept. of Astronomy, University of Wisconsin, Madison, WI 53706, USA}
\affiliation{Dept. of Physics and Wisconsin IceCube Particle Astrophysics Center, University of Wisconsin, Madison, WI 53706, USA}
\affiliation{Institute of Physics, University of Mainz, Staudinger Weg 7, D-55099 Mainz, Germany}
\affiliation{Department of Physics, Marquette University, Milwaukee, WI, 53201, USA}
\affiliation{Institut f{\"u}r Kernphysik, Westf{\"a}lische Wilhelms-Universit{\"a}t M{\"u}nster, D-48149 M{\"u}nster, Germany}
\affiliation{Bartol Research Institute and Dept. of Physics and Astronomy, University of Delaware, Newark, DE 19716, USA}
\affiliation{Dept. of Physics, Yale University, New Haven, CT 06520, USA}
\affiliation{Dept. of Physics, University of Oxford, Parks Road, Oxford OX1 3PU, UK}
\affiliation{Dept. of Physics, Drexel University, 3141 Chestnut Street, Philadelphia, PA 19104, USA}
\affiliation{Physics Department, South Dakota School of Mines and Technology, Rapid City, SD 57701, USA}
\affiliation{Dept. of Physics, University of Wisconsin, River Falls, WI 54022, USA}
\affiliation{Dept. of Physics and Astronomy, University of Rochester, Rochester, NY 14627, USA}
\affiliation{Oskar Klein Centre and Dept. of Physics, Stockholm University, SE-10691 Stockholm, Sweden}
\affiliation{Dept. of Physics and Astronomy, Stony Brook University, Stony Brook, NY 11794-3800, USA}
\affiliation{Dept. of Physics, Sungkyunkwan University, Suwon 16419, Korea}
\affiliation{Institute of Basic Science, Sungkyunkwan University, Suwon 16419, Korea}
\affiliation{Dept. of Physics and Astronomy, University of Alabama, Tuscaloosa, AL 35487, USA}
\affiliation{Dept. of Astronomy and Astrophysics, Pennsylvania State University, University Park, PA 16802, USA}
\affiliation{Dept. of Physics, Pennsylvania State University, University Park, PA 16802, USA}
\affiliation{Dept. of Physics and Astronomy, Uppsala University, Box 516, S-75120 Uppsala, Sweden}
\affiliation{Dept. of Physics, University of Wuppertal, D-42119 Wuppertal, Germany}
\affiliation{DESY, D-15738 Zeuthen, Germany}
\affiliation{Department of Physics, Columbia University in the City of New York, NY 10027, USA}
\affiliation{Deparment of Physics, University of Florida, Gainesville, FL 32611-8440, USA}

\author{M. G. Aartsen}
\affiliation{Dept. of Physics and Astronomy, University of Canterbury, Private Bag 4800, Christchurch, New Zealand}
\author{M. Ackermann}
\affiliation{DESY, D-15738 Zeuthen, Germany}
\author{J. Adams}
\affiliation{Dept. of Physics and Astronomy, University of Canterbury, Private Bag 4800, Christchurch, New Zealand}
\author{J. A. Aguilar}
\affiliation{Universit{\'e} Libre de Bruxelles, Science Faculty CP230, B-1050 Brussels, Belgium}
\author{M. Ahlers}
\affiliation{Niels Bohr Institute, University of Copenhagen, DK-2100 Copenhagen, Denmark}
\author{M. Ahrens}
\affiliation{Oskar Klein Centre and Dept. of Physics, Stockholm University, SE-10691 Stockholm, Sweden}
\author{C. Alispach}
\affiliation{D{\'e}partement de physique nucl{\'e}aire et corpusculaire, Universit{\'e} de Gen{\`e}ve, CH-1211 Gen{\`e}ve, Switzerland}
\author{K. Andeen}
\affiliation{Department of Physics, Marquette University, Milwaukee, WI, 53201, USA}
\author{T. Anderson}
\affiliation{Dept. of Physics, Pennsylvania State University, University Park, PA 16802, USA}
\author{I. Ansseau}
\affiliation{Universit{\'e} Libre de Bruxelles, Science Faculty CP230, B-1050 Brussels, Belgium}
% ...

\begin{abstract}

Using the IceCube Neutrino Observatory, we search for high-energy neutrino emission coincident with compact binary mergers observed by the LIGO and Virgo gravitational wave (GW) detectors during their first and second observing runs. We present results from two searches targeting emission coincident with the sky localization of each gravitational wave event within a 1000 second time window centered around the reported merger time. One search uses a model-independent unbinned maximum likelihood analysis, which uses neutrino data from IceCube to search for point-like neutrino sources consistent with the sky localization of GW events. The other uses the Low-Latency Algorithm for Multi-messenger Astrophysics, which incorporates astrophysical priors through a Bayesian framework and includes LIGO-Virgo detector characteristics to determine the association between the GW source and the neutrinos. No significant neutrino coincidence is seen by either search during the first two observing runs of the LIGO-Virgo detectors. We set upper limits on the time-integrated neutrino emission within the 1000 second window for each of the 11 GW events. These limits range from 0.02-0.7 $\mathrm{GeV~cm^{-2}}$. We also set limits on the total isotropic equivalent energy, $E_{\mathrm{iso}}$, emitted in high-energy neutrinos by each GW event. These limits range from 1.7 $\times$ 10$^{51}$ - 1.8 $\times$ 10$^{55}$ erg. We conclude with an outlook for LIGO-Virgo observing run O3, during which both analyses are running in real time.

\end{abstract}

\section{Introduction} \label{sec:intro}

The discovery of gravitational waves (GWs)~\citep{PhysRevLett.116.061102} and the discovery of astrophysical high-energy neutrinos has opened new opportunities for astrophysics with the possibility of finding multi-messenger fingerprints~\citep{2008CQGra..25k4051A,GBM:2017lvd,ic1709022mm}.

Both GW and high-energy neutrino detections are now reported on a weekly to monthly basis \citep{gracedb,amon}. All GW detections by the LIGO and Virgo observatories published to date are well understood in the context of coalescing compact binary systems~\citep{TheLIGOScientific:2014jea,TheVirgo:2014hva,Abbott_2019}, while the dominant origin of the detected astrophysical neutrinos is less certain \citep{Aartsen:2019fau}. The joint observation of GWs and high-energy neutrinos could shed light on the underlying source population responsible for neutrino emission, and would give insight into the connection between the interaction of compact objects and the properties of energetic outflows driven by their interactions \citep{2013RvMP...85.1401A,2013CQGra..30l3001B}. It could also shed light on astrophysical sources which are electromagnetically obscured due to a surrounding dense environment, but are transparent to neutrinos \citep{2003PhRvD..68h3001R,2012PhRvD..86h3007B,Kimura:2018vvz}.

For over a decade there has been a sustained effort dedicated to discovering joint sources of GWs and high-energy neutrinos~\citep{ 2008CQGra..25k4039A, 2009IJMPD..18.1655V, 2011PhRvL.107y1101B, 2012PhRvD..85j3004B, 2013JCAP...06..008A, 2013RvMP...85.1401A, 2014PhRvD..90j2002A, 2016PhRvD..93l2010A, 2017ApJ...850L..35A,2017PhRvD..96b2005A}. The first observational constraints on common sources were obtained from non-detections \citep{2011PhRvL.107y1101B}. Initial LIGO-Virgo data were analyzed in two independent searches, one using ANTARES data \citep{2013JCAP...06..008A} and the other using IceCube \citep{2014PhRvD..90j2002A} data. After the onset of the era of advanced GW detectors, GW discoveries led to an increased interest in finding neutrino counterparts of GWs~\citep{2016PhRvD..93l2010A,2017PhRvD..96b2005A,2017ApJ...850L..35A,2018JETPL.108..787A,2018ApJ...857L...4A,2018JETPL.107..398P}. Additionally, a subthreshold search looked for events where neither the GW nor the neutrino trigger could be independently confirmed to be astrophysical \citep{Albert_2019}. No statistically significant correlations were found in these searches.

The near realtime availability of GW and high-energy neutrino data enables low-latency joint searches. This allows for follow-up observations within a few minutes of an alert which is crucial for observing prompt emission with multiple messengers. The current Low-Latency Algorithm for Multi-messenger Astrophysics (LLAMA) evolved from the previous joint GW-neutrino search pipeline~\citep{2012PhRvD..85j3004B} that produced the bulk of the published results up to date. During the second LIGO-Virgo observing run (O2), LLAMA combined data from LIGO, Virgo, and IceCube, and disseminated results to electromagnetic follow-up partners \citep{2019arXiv190105486C}. The application of LLAMA to the data from the O2 observing run provided the proof-of-concept for a reliable low-latency multi-messenger pipeline. LLAMA with an improved significance measure \citep{PhysRevD.100.083017}  continues to run during the current third observing run \citep{2019arXiv190105486C,Keivani:2019smf}, together with the online version of the analysis using the maximum likelihood method \citep{Hussain:2019xzb} discussed below.

\subsection{First and second observing runs of the advanced gravitational wave detector network}

During the first and second observing runs of the advanced gravitational wave detector network \citep{TheLIGOScientific:2014jea,TheVirgo:2014hva}, the LIGO and Virgo Collaborations (LVC) searched the collected data and discovered signals from compact binary mergers. Three binary black hole (BBH) mergers were detected during O1 (September 12th, 2015 to January 19th, 2016), and an additional seven BBH events were detected during O2 (November 30th, 2016 to August 25th, 2017). The LVC also discovered the very first binary neutron star (BNS) inspiral signal during O2, on August 17th, 2017 (GW170817). GWTC-1, a Gravitational wave Transient Catalog of compact binary mergers observed by LIGO and Virgo during the first and second observing runs, provides information on source properties and localization on the aforementioned eleven GW events \citep{Abbott_2019}.

The observed masses of BBH components span a wide range from \ensuremath{7.7_{-2.5}^{+2.2}}~\Msun to \ensuremath{50.2_{-10.2}^{+16.2}}~\Msun, while for the BNS event, GW170817, the NS masses are \ensuremath{1.27_{-0.09}^{+0.09}}~\Msun and \ensuremath{1.46_{-0.10}^{+0.12}}~\Msun. The O1 and O2 gravitational-wave events range in distance between \ensuremath{40_{-15}^{+8}}Mpc for GW170817 to \ensuremath{2840_{-1360}^{+1400}}Mpc for GW170729 \citep{Abbott_2019}. All references to distance refer to the luminosity distance, $D_L$.

The time difference between the signals observed at the different interferometers in the GW detector network, together with the phase and amplitude of the detected GW events, enables sky localization of the source~\citep{2016PhRvD..93b4013S,2015PhRvD..91d2003V}. The 90\% credible regions of the sky localizations for the GWTC-1 events span 16 to 1666 square degrees. Most events were observed only with the two LIGO detectors, thus the corresponding sky localization regions for these events are rather large. The presence of the Virgo detector, especially for closer and/or higher-mass events, significantly improves the localization (see Table \ref{tab:results}), though in all cases the GW localization is poor with respect to the angular uncertainty of high-energy neutrino events which tend to be less than a few square degrees.

IceCube and LIGO-Virgo have previously reported on the search for high-energy neutrino counterparts for four of the eleven GWTC-1 catalog events (GW150914, GW151012, and GW151226 from O1 and GW170817 from O2)~\citep{2016PhRvD..93l2010A,2017PhRvD..96b2005A,2017ApJ...850L..35A}. Moreover, during O2 the LLAMA disseminated joint search results \citep{2019arXiv190105486C} to electromagnetic follow-up partners for six events (GW170104, GW170608, GW170809, GW170814, GW170817, and GW170823) as well as for additional candidate events that were distributed to LVC partners~\citep{2019ApJ...875..161A}. In this paper we present the results for all 11 GW events in the GWTC-1 Catalog from a model-independent unbinned maximum likelihood search as well as an updated version of the previous LLAMA search. 

Both searches presented below search for prompt neutrino emission within a 1000~s time window centered around the GW event time. This search window is chosen based on a range of neutrino emission mechanisms from gamma-ray bursts and is a conservative estimate of the difference in arrival times of the GW and neutrinos \citep{Baret:2011tk}. Searches for longer-timescale neutrino emission are in development but are not considered here.

\subsection{IceCube}
IceCube is a cubic-kilometer neutrino observatory located at the geographic South Pole \citep{Aartsen:2016nxy}. IceCube consists of 86 strings, each of which holds 60 Digital Optical Modules (DOMs) which are located at depths between 1.5 km and 2.5 km in the Antarctic ice. These DOMs contain photomultiplier tubes which detect the Cherenkov light radiated by charged secondaries of neutrino charged-current and neutral-current interactions. 

\begin{figure}[h!]
    \centering
    \includegraphics[width=0.65\textwidth]{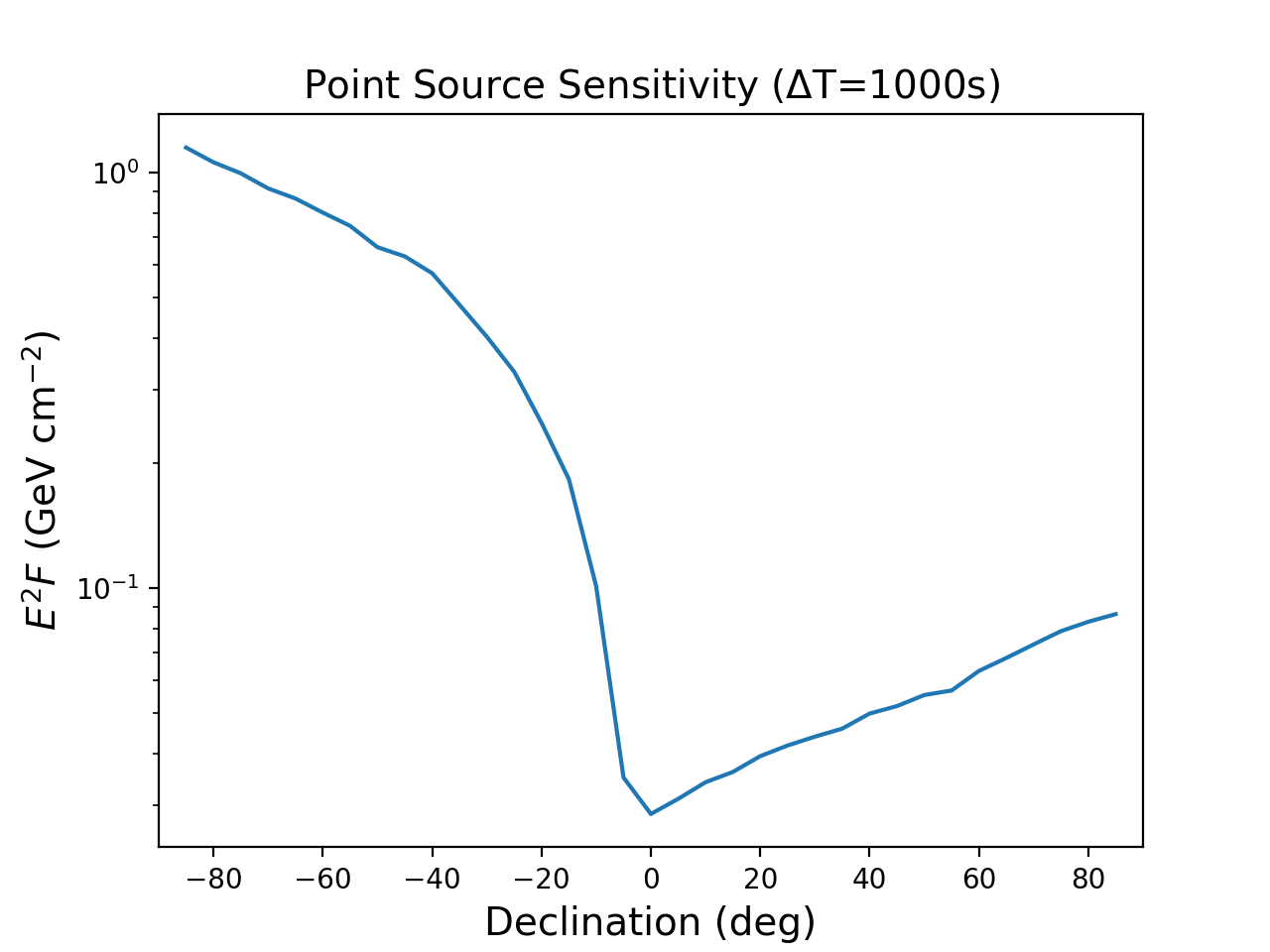}
    \caption{IceCube's 90\% C.L. sensitivity to a transient point source within 1000~s assuming an $E^{-2}$ spectrum. We assume uniform emission within the time window. Here, $E^2F$ is the energy scaled, time-integrated flux, where the time-integrated flux is defined as $F = dN/dE~dA$. The choice of an $E^{-2}$ flux provides an optimistic limit on the sensitivity of IceCube and is motivated by Fermi acceleration. }
    \label{fig:ps_sens}
\end{figure}

IceCube's sensitivity to a point source is strongly dependent on the declination of the source. Figure \ref{fig:ps_sens} shows IceCube's sensitivity to a transient point source in a 1000~s time window as a function of declination. The predominant background in the Northern Sky consists of atmospheric neutrinos which can travel large column depths through the earth, while any atmospheric muons will be absorbed in the earth \citep{Formaggio:2013kya}. This is not the case in the Southern Sky, where atmospheric muons have enough energy to travel from the atmosphere to the detector resulting in a much higher total background rate. Therefore, IceCube is much more sensitive to sources in the Northern Sky.

IceCube's nearly 100\% uptime and continuous 4$\pi$ steradian field of view make it an ideal observatory for multi-messenger programs, both to trigger other observatories as well as perform follow-ups \citep{Aartsen:2016lmt}. In the case of multi-messenger follow-ups of GW events, IceCube is able to search for neutrinos within the full GW localization region for all reported GW events.

Both analyses described in this paper use the same neutrino data which come from the IceCube Gamma-ray Follow Up (GFU) sample \citep{Kintscher:2016uqh} but use different statistical methods and test different hypotheses. This sample is used for low-latency as well as offline analyses in IceCube. It consists of through-going muon tracks primarily induced by cosmic-ray interactions in the atmosphere. These backgrounds consist of downgoing muons as well as upgoing muons from atmospheric muon-neutrino interactions. The rate of background events in the sample is roughly three orders of magnitude larger than the rate of astrophysical neutrinos. Overall, the sample has a 6.7 mHz all-sky event rate and the events in the sample have a median angular resolution of $\lesssim$1$^\circ$ for energies above 1 TeV. The GFU sample is ideal for realtime multi-messenger follow ups due to its low latency ($\sim$30~s) and good angular resolution. 

\section{Methods}{\label{sec:methods}}
\subsection{Unbinned Maximum Likelihood}\label{sec:maxLLH}
The unbinned maximum likelihood analysis tests for a point-like neutrino source being consistent with the localization of the GW source detected by LVC. The method uses the GW skymap as a spatial prior in a neutrino point-source likelihood \citep{Schumacher:2019qdx}.

The sky is divided into equal-area bins using HEALPix \citep{healpix}. The pixels are roughly 0.01 deg$^2$, which is about the order of magnitude of the best angular resolution of the neutrino events in the GFU sample. In each pixel, the likelihood is maximized with respect to the number of signal neutrinos, $n_s$, and the source spectral index, $\gamma$. This procedure yields a test statistic (TS) in each pixel of the form

\begin{equation}
    \mathrm{TS} = 2 \ln\left(\frac{\textit{L}(\hat{n_{\mathrm{s}}},\hat{\gamma})}{\textit{L}(n_{\mathrm{s}}=0)} \right)
\end{equation}

\noindent where $\hat{n_{\mathrm{s}}}$ and $\hat{\gamma}$ are the values of $n_{\mathrm{s}}$ and $\gamma$ which maximize the likelihood, \textit{L}.

The resulting test statistics are then weighted by the spatial prior which is a penalty at every pixel derived from the probability distribution of the GW event over the sky. This yields a weighted test statistic 

\begin{equation}
    \Lambda = 2 \ln\left(\frac{\textit{L}(\hat{n_{\mathrm{s}}},\hat{\gamma}) \cdot w}{\textit{L}(n_{\mathrm{s}}=0)} \right) = \mathrm{TS} + 2\ln(w)
\end{equation}

\noindent where \textit{w} is the weight derived from the GW localization, $w$ = $P_{\mathrm{GW}}(\Omega)/A_{pix}$, where $P_{GW}(\Omega)$ is the GW localization probability as a function of position on the sky and $A_{pix}$ is the area of the pixels on the sky. A full, detailed description of this method can be found in \cite{Hussain:2019xzb}.

\begin{figure}[h!]
    \centering
    \includegraphics[width=0.65\textwidth]{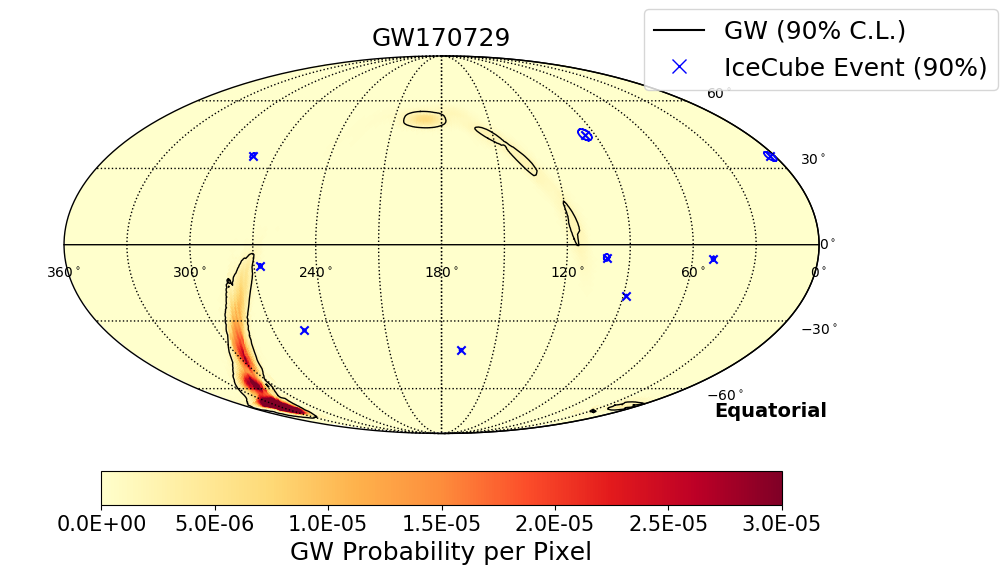}
    \caption{An example of a neutrino follow up to GW170729 \citep{Abbott_2019}, a binary black hole merger (BBH) during the O2 Observing run. Here, the neutrinos observed within $\pm500$~s of the GW event are represented by blue crosses with 90\% containment angular errors and the GW localization PDF is shown in red with the 90\% credible region in black. Note that some of the 90\% containment regions for the neutrinos are smaller than the blue crosses so the error regions aren't visible.}
    \label{fig:skymap}
\end{figure}

For a given GW event, we test for coincident neutrinos by considering a $\pm$500~s time window centered on the GW event time. To keep the analysis model independent, neutrinos are assumed to be emitted uniformly within the $\pm$ 500~s time window. An example of a GW event overlaid with IceCube neutrinos within $\pm$500~s of the GW event is shown in  Figure \ref{fig:skymap}. For a gallery of all 11 skymaps with overlaid neutrinos, see Figure \ref{fig:skymap_gallery} in the Supplement.

We quantify the significance of a given observation by comparing our maximum observed $\Lambda$ over the full sky to a background-only distribution and computing the resulting p-value. The background distribution is built from 30,000 trials using randomized neutrino events from the data themselves. To randomize the neutrino events used in each trial, the event arrival times are randomly shuffled while keeping the local coordinates of the events. This procedure preserves the time structure of the data set while assigning a random right ascension to each neutrino event, thus producing a unique sample for each trial. An example of a background-only test statistic distribution for GW170729 is shown in Figure \ref{fig:bkgTS} in the Supplement. We fix the GW skymap for these trials so each GW event in the catalog has a unique background distribution. The associated p-value for a given GW event quantifies the chance probability of a set of background neutrinos having a significant correlation with the localization of the GW event. 30,000 trials per GW event yield enough statistics to compute accurate p-values while still being computationally feasible.

\subsection{Low-Latency Algorithm for Multi-messenger Astrophysics (LLAMA)}

The LLAMA search calculates the odds ratio of having a multi-messenger counterpart for a GW event versus the GW originating from noise or not having a counterpart (see \cite{2019arXiv190105486C,Keivani:2019smf} for more information). This odds ratio is used as the test statistic (TS) for the search. The calculation of the odds ratio is based on a Bayesian framework by assuming a model for the multi-messenger sources  \citep{PhysRevD.100.083017}. The method uses a distribution for the astrophysical high-energy neutrinos' total emission energies, a spectrum for individual neutrino's differential energy density, IceCube's detector response, and the spatial position reconstruction of the GW detection to estimate the expected number of neutrinos to be detected from the GW event. More neutrinos are expected to be detected from closer and otherwise identical events and thus closer events are favored. A log-uniform distribution between \(10^{46}-10^{51}\) ergs is used for the distribution of the total high-energy neutrino emission energy. The assumed neutrino emission spectrum is a power law with exponent \(\gamma=-2\). The method's input parameters are the detection times and localizations of candidate neutrinos and the GW, the reconstructed energies of the candidate neutrinos, the reconstructed distance of the GW, and the signal to noise ratio (SNR) of the GW. For GW events which have confirmed detections, as is the case with the 11 events considered here, the probability of the GW event arising from background is assumed to be 0, and the SNR information from the GW is not used. The method assumes neutrinos and GWs are emitted uniformly in \(\pm 250\) s around the joint astrophysical event time \citep{Baret:2011tk}. This results in a triangular distribution for the time difference between neutrinos and GWs with a maximum of \(\pm 500\) s time difference, where neutrinos temporally closer to the GW are favored. The p-value for each event is found by comparing the observed TS value for the event to a background distribution. The background distribution is built by running the analysis on injected GWs which are distributed uniformly in the comoving volume and detected with the sensitivity of the GW detectors during O1 and O2 runs, which have similar sensitivities. The background distributions for BBH and BNS events are kept separate due to the significantly different distance distributions of their detections. Figure \ref{fig:llama_bg_dist} in the Supplement shows the background TS distribution for BBH mergers detected with 3 GW detectors, aLIGO Hanford, aLIGO Livingston, and AdVirgo.

\section{Results}
No significant neutrino correlation was found for the 11 GWs in the O1 and O2 observing runs. We set upper limits (ULs) on the energy scaled time-integrated flux, $E^2F$, assuming an $E^{-2}$ source spectrum. A single UL on $E^2F$ is derived for each GW event. This is done via signal injection trials which are described in detail in the supplement. We also report a range of upper limits for each GW event. This range corresponds to the minimum and maximum UL that IceCube can set within the 90\% credible region of the GW event. The results are shown in Table \ref{tab:results}. 

We also compute an upper limit on an astrophysical quantity of interest, the isotropic equivalent energy, $E_{\mathrm{iso}}$, assuming an $E^{-2}$ source spectrum. Physically, this quantity represents the total energy emitted in neutrinos during the 1000~s time window, assuming spherically symmetric emission. $E_{\mathrm{iso}}$ encodes the intrinsic energy emitted by the source combined with the relative Doppler boost of the emission. Due to the uncertainties in the 3D localization of the GW source, we also marginalize over the 3D position of the source to get a marginalized UL on $E_{\mathrm{iso}}$. This results in a single UL which encapsulates the significant uncertainties in the position of the source. 

We relate $E_{\mathrm{iso}}$ to the neutrino flux observed by IceCube:

\begin{equation}
\begin{split}
\frac{E_{\mathrm{iso}}}{4\pi r^2} & = \int_{E_1}^{E_2} \Phi(E) E \Delta t dE \\
 & = \Phi_0 \Delta t\ E_0^2 \ln\left(\frac{E_2}{E_1}\right),
\end{split}
\end{equation}

\noindent where we assume the flux, $\Phi(E)$, follows a simple power-law spectrum, $\Phi(E) = \Phi_0~(E/E_0)^{-2}$. The flux normalization $\Phi_0$ has units of $\mathrm{GeV^{-1}~cm^{-2}~s^{-1}}$. The power-law flux is related to the flux defined in Figure \ref{fig:ps_sens} as $E^2F$ = $E^2\Phi (E) \Delta t$. $E_1$ and $E_2$ are the minimum and maximum energy of the neutrino Monte Carlo sample respectively, where $E_1$=10~GeV and $E_2$=$10^{9.5}$~GeV. Thus, our $E_{\mathrm{iso}}$ ULs are to be taken as limits on the energy emitted in neutrinos between $E_1$ and $E_2$.

Next we relate $E_{\mathrm{iso}}$ to the expected number of neutrino events, $\mu$, observed at IceCube. To do this, we need to take into account IceCube's effective area, $A_\mathrm{eff}(E,\delta)$, which is strongly dependent on declination, and we also marginalize over the 3D position of the source:

\begin{equation}
    \begin{split}
        \mu & = \int A_\mathrm{eff}(E,\delta) \Phi(E) P(\Omega,\mathrm{r})\Delta t\ dE dV \\
            & = \sum_{i=0}^{N} p(\Omega_i) \int A_\mathrm{eff}(E,\delta_i) \frac{E_{\mathrm{iso}}E^{-2}}{4 \pi \ln\left(\frac{E_2}{E_1}\right)}\ dE \int_0^{\infty} \frac{1}{r^2} p(r|\Omega_i)\ dr ,
    \end{split}
\end{equation}

\noindent where the summation is over the pixels in the skymap of the GW event, $P(\Omega,\mathrm{r})$ is the 3D location PDF of the GW source, and $p(\Omega_i)$ is the probability per pixel. Since $p(\Omega_i)$ is a discrete quantity, we transform the integral into a sum over the pixels in the sky. The quantity $p(r|\Omega_i)$ is the per-pixel luminosity distribution of the form 

\begin{equation}
    p(r|\Omega_i) = \frac{r^2 d_{\mathrm{norm},i}}{\sqrt{2\pi d_{\sigma,i}^2}} \exp{\left[\frac{-(r-d_{\mu,i})^2}{2d_{\sigma,i}^2}\right]} ,
\end{equation}

\noindent where $d_{\mathrm{norm},i}$, $d_{\mu,i}$, and $d_{\sigma,i}$ are parameters fitted in the $i^{th}$ pixel to create a per-pixel distance distribution. For details on LIGO-Virgo skymaps and localizations see \cite{Singer:2016erz}.

Using these distributions we can numerically solve for the expected number of neutrino events at IceCube after marginalizing the localization uncertainty. We then perform trials with additional signal injection to compute a 90\% UL on $E_{\mathrm{iso}}$. The results for this calculation are shown in Table \ref{tab:results}. This $E_{\mathrm{iso}}$ limit was only computed with the maximum likelihood analysis described in Section \ref{sec:maxLLH}. See \cite{veske2020neutrino} for LLAMA search's $E_{\mathrm{iso}}$ limits for the first 3 GW events with a different statistical treatment.

\begin{table}[h!]
\begin{center}
\footnotesize
\setlength\tabcolsep{4pt}
\begin{tabular}{|c|c|c|c|c|c|c|c|c|c|c|}
 \hline
 \multicolumn{11}{|c|}{O1 and O2 Detections} \\ 
 \hline
 \multicolumn{6}{|c|}{} & \multicolumn{3}{|c|}{Maximum Likelihood} & \multicolumn{2}{|c|}{LLAMA} \\
 \hline
 \multirow{2}{*}{Event} & \multirow{2}{*}{Type}  & \multirow{2}{*}{Detectors} & $\Omega$ & D$_L$ & UL Range  & \multirow{2}{*}{p-value} & UL   & $E_{\mathrm{iso}}$ UL & \multirow{2}{*}{p-value} & UL  \\
 & & & ($\mathrm{deg}^2$) & (Mpc) & (GeVcm$^{-2}$) & & (GeVcm$^{-2}$) & (erg) & & (GeVcm$^{-2}$) \\
 \hline
 GW150914   & BBH & LH  & 182   &  \ensuremath{440_{-170}^{+150}}  &  0.0296 - 1.03   & 0.51 &  0.66  & 5.10  $\times$  10$^{53}$ &   0.29    & 0.70 \\ \hline 
 GW151012   & BBH & LH  & 1523  &  \ensuremath{1080_{-490}^{+550}} &  0.0286 - 0.821  & 0.83 &  0.16  & 7.50  $\times$  10$^{53}$ &     0.82  & 0.18 \\ \hline 
 GW151226   & BBH & LH  & 1033  &  \ensuremath{450_{-190}^{+180}}  &  0.0286 - 0.904  & 0.74 &  0.22  & 1.74  $\times$  10$^{53}$ &    0.26   & 0.21 \\ \hline 
 GW170104   & BBH & LH  & 921   &  \ensuremath{990_{-430}^{+440}}  &  0.0286 - 0.667  & 0.54 &  0.044 & 1.81  $\times$  10$^{53}$ & 0.16  & 0.055 \\ \hline
 GW170608   & BBH & LH  & 392   &  \ensuremath{320_{-110}^{+120}}  &  0.0309 - 0.0821 & 0.61 &  0.037 & 1.37  $\times$  10$^{52}$ & 0.97  & 0.038 \\ \hline
 GW170729   & BBH & LH  & 1041  &  \ensuremath{2840_{-1360}^{+1400}} &  0.0286 - 1.02   & 0.21 &  0.62  & 1.80  $\times$  10$^{55}$ & 0.17  & 0.62 \\ \hline 
 GW170809   & BBH & LH  & 308   &  \ensuremath{1030_{-390}^{+320}} &  0.0568 - 0.758  & 0.60 &  0.27  & 1.02  $\times$  10$^{54}$ & 0.83  & 0.26 \\ \hline 
 GW170814   & BBH & LHV & 87    &  \ensuremath{600_{-220}^{+150}}  &  0.488  - 0.711  & 0.83 &  0.45  & 5.47  $\times$  10$^{53}$ & 1.0   & 0.43 \\ \hline
 GW170817   & BNS & LHV & 16    &  \ensuremath{40_{-15}^{+7}}   &  0.180  - 0.429  & 0.19 &  0.27  & 1.67  $\times$  10$^{51}$ & 0.94  & 0.25 \\ \hline 
 GW170818   & BBH & LHV & 39    &  \ensuremath{1060_{-380}^{+420}} &  0.0364 - 0.0431 & 0.58 &  0.028 & 1.17  $\times$  10$^{53}$ & 0.40  & 0.028 \\ \hline
 GW170823   & BBH & LH  & 1666  &  \ensuremath{1940_{-900}^{+970}} &  0.0286 - 0.796  & 0.75 &  0.18  & 2.33  $\times$  10$^{54}$ & 0.25  & 0.18 \\ \hline
\end{tabular}
\end{center}
\caption{Results for every detected GW from the O1 and O2 observing runs. Here $\Omega$ is the area of the 90\% credible region of the GW and D$_L$ is the reported median luminosity distance. These values are taken from GWTC-1 \citep{Abbott_2019}. The Detectors column indicates which of the three LIGO-Virgo detectors detected the GW.  We also report 90\% ULs on the energy scaled time-integrated flux, $E^2F$, from both analyses. The UL Range column shows the minimum and maximum 90\% ULs assuming a point source hypothesis within the 90\% credible region of the GW skymap. $E_{\mathrm{iso}}$ is the UL on the isotropic equivalent energy emitted in neutrinos during 1000~s. Note that error bars on derived energy quantities are not shown for clarity but are dominated by the error in the distance measurements of each GW. BBH = Binary Black Hole, BNS = Binary Neutron Star.}
\label{tab:results}
\end{table}

\begin{figure}[h!]
    \centering
    \includegraphics[width=0.75\textwidth]{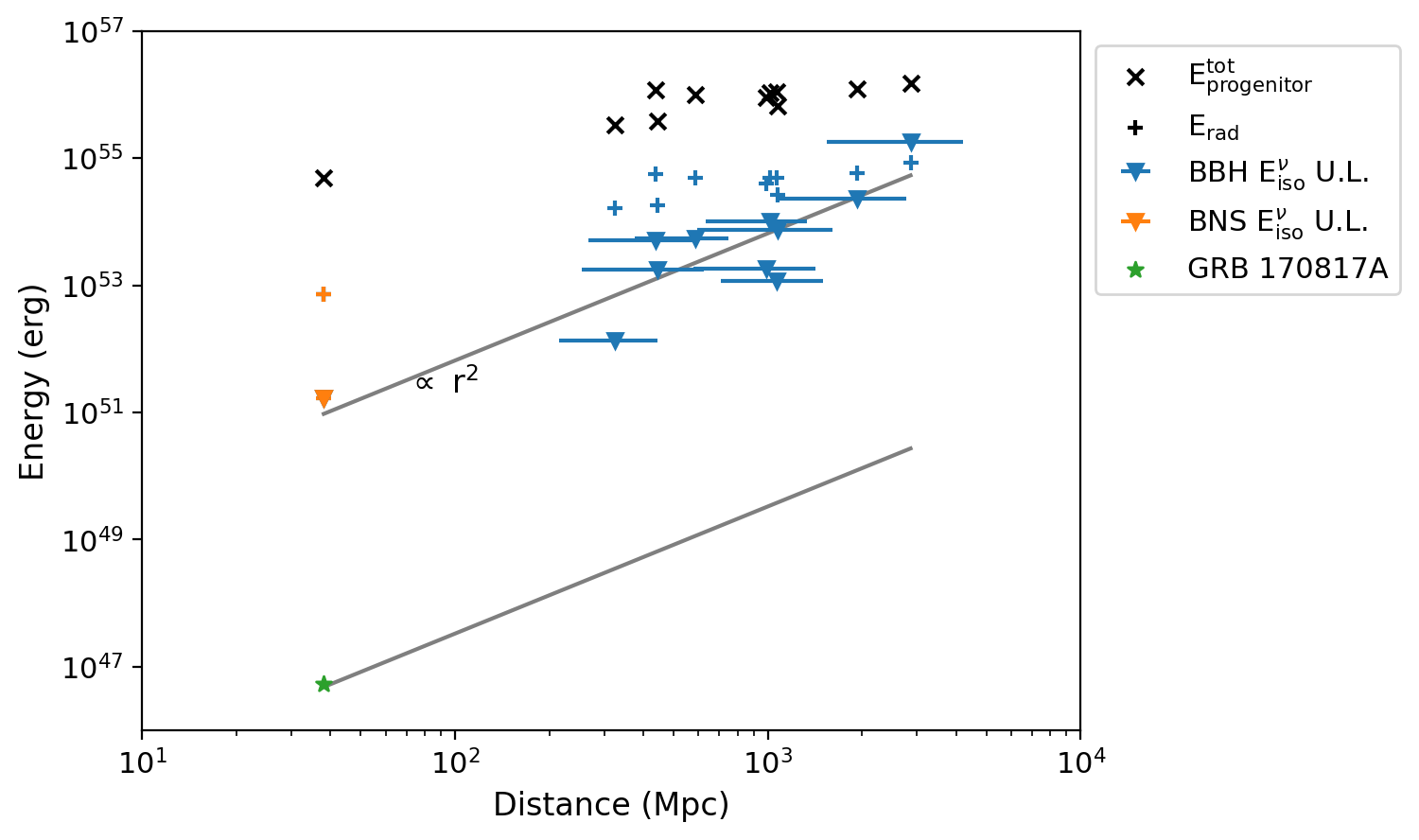}
    \caption{90\% UL on the  isotropic equivalent energy emitted in neutrinos during a 1000~s time window (blue and orange triangles). The $E_{\mathrm{iso}}$ ULs are fit to an $r^2$ scaling which is motivated by geometric arguments. $E_{\mathrm{progenitor}}^{\mathrm{tot}}$ (black cross) is the total rest mass energy of the progenitors and $E_{\mathrm{rad}}$ (orange/blue plus) is the total radiated energy of the binary system. While not all of the progenitor energy is available for acceleration processes, we show it here as a relevant energy scale in the binary system. The distance and 90\% credible intervals are taken from the first GW catalog \citep{Abbott_2019}. The distance errors for GW170817 are much smaller than the BBHs because of the precise measurements of the host galaxy \citep{Cantiello:2018ffy}. Note that the distance error bars also apply to the $E_{\mathrm{rad}}$ and $E_{\mathrm{progenitor}}^{\mathrm{tot}}$ measurements but are not shown here for clarity. Shown in green is the measured $E_{\mathrm{iso}}$ for GRB 170817A by Fermi GBM taken from \cite{Monitor:2017mdv}. The assumed \textit{r}$^2$ scaling is fit to this measured value.}
    \label{fig:eiso}
\end{figure}

Figure \ref{fig:eiso} shows the upper limits on $E_{\mathrm{iso}}$ as a function of the mean distance to the source marginalized over the sky. We find that $E_{\mathrm{iso}}$ scales roughly with the square of the distance which is motivated from geometrical arguments. There is significant scatter around the $r^2$ scaling due to IceCube detector effects and the significant uncertainties in the GW source positions. Also shown in Figure \ref{fig:eiso} is the total radiated energy for each GW as well as the total rest mass energy of the initial binary system. Our $E_{\mathrm{iso}}$ ULs show that the UL on the total energy radiated in neutrinos within the 1000~s time window is up to 2 orders of magnitude smaller than the total radiated energy of the binary system, while for some events, it is about the same order of magnitude.

\section{Conclusion}
Advances in detector technologies used in GW, electromagnetic, and neutrino astronomy have led to a scientific revolution, rapidly expanding the fields' cosmic and scientific horizons. Finding evidence of a joint GW-neutrino source would further expand our understanding of the sources of high-energy neutrinos and the energetic outflows driven by the interaction of compact objects.

We performed two different searches for neutrino emission from the 11 compact binary mergers in the first gravitational wave transients catalog, GWTC-1, from LIGO-Virgo. We found no significant neutrino emission from any of the 11 GW events and therefore placed upper limits on the energy-scaled time-integrated neutrino flux (see Table \ref{tab:results}). 
In addition to upper limits on neutrino flux, we placed upper limits on the isotropic equivalent energy emitted in neutrinos, $E_{\mathrm{iso}}$, during the 1000~s time window (see Table \ref{tab:results}).  These limits show that BBH mergers emit up to 2 orders of magnitude less energy in high-energy neutrinos than they do in gravitational waves.
For the one BNS merger, GW170817, the UL on the energy emitted in neutrinos is about 2 orders of magnitude lower than the energy radiated in gravitational waves. We also compare to measurements of the energy emitted in gamma-rays of the associated short gamma-ray burst, GRB 170817A. Fermi GBM reported an upper limit on $E_{\mathrm{iso}}$ which is over six orders of magnitude lower than the energy radiated in gravitational waves and over 4 orders of magnitude lower than the UL on the energy emitted in neutrinos. \citep{Monitor:2017mdv}.

In addition to searching for neutrino emission from the 11 mergers in GWTC-1, there are two pipelines implementing the two methods described in Section \ref{sec:methods} in low-latency during the O3 observing run. These low-latency searches are particularly useful in informing electromagnetic observatories where to point to search for optical counterparts. These analyses are described in detail in \cite{Hussain:2019xzb} and \cite{2019arXiv190105486C}, and will the subject of a future publication.

Searches for neutrino emission from longer time windows are also ongoing. These searches target neutrino emission from binary neutron star or neutron star-black hole mergers on a two-week timescale. Neutrinos from kilonovae or ejected material from mergers involving neutron stars are potential sources of high-energy neutrinos for weeks after the initial merger \citep{Kimura:2018vvz, Fang:2017tla}. Our sample of potential joint GW and neutrino sources continues to grow as more compact binary mergers are detected. With the $5\times$ to $6\times$ higher statistics expected in O3 \citep{2018LRR....21....3A}, we can search for a possible underlying population of joint sources.

\section*{Acknowledgements}
USA {\textendash} U.S. National Science Foundation-Office of Polar Programs,
U.S. National Science Foundation-Physics Division,
Wisconsin Alumni Research Foundation,
Center for High Throughput Computing (CHTC) at the University of Wisconsin-Madison,
Open Science Grid (OSG),
Extreme Science and Engineering Discovery Environment (XSEDE),
U.S. Department of Energy-National Energy Research Scientific Computing Center,
Particle astrophysics research computing center at the University of Maryland,
Institute for Cyber-Enabled Research at Michigan State University,
and Astroparticle physics computational facility at Marquette University;
Belgium {\textendash} Funds for Scientific Research (FRS-FNRS and FWO),
FWO Odysseus and Big Science programmes,
and Belgian Federal Science Policy Office (Belspo);
Germany {\textendash} Bundesministerium f{\"u}r Bildung und Forschung (BMBF),
Deutsche Forschungsgemeinschaft (DFG),
Helmholtz Alliance for Astroparticle Physics (HAP),
Initiative and Networking Fund of the Helmholtz Association,
Deutsches Elektronen Synchrotron (DESY),
and High Performance Computing cluster of the RWTH Aachen;
Sweden {\textendash} Swedish Research Council,
Swedish Polar Research Secretariat,
Swedish National Infrastructure for Computing (SNIC),
and Knut and Alice Wallenberg Foundation;
Australia {\textendash} Australian Research Council;
Canada {\textendash} Natural Sciences and Engineering Research Council of Canada,
Calcul Qu{\'e}bec, Compute Ontario, Canada Foundation for Innovation, WestGrid, and Compute Canada;
Denmark {\textendash} Villum Fonden, Danish National Research Foundation (DNRF), Carlsberg Foundation;
New Zealand {\textendash} Marsden Fund;
Japan {\textendash} Japan Society for Promotion of Science (JSPS)
and Institute for Global Prominent Research (IGPR) of Chiba University;
Korea {\textendash} National Research Foundation of Korea (NRF);
Switzerland {\textendash} Swiss National Science Foundation (SNSF);
United Kingdom {\textendash} Department of Physics, University of Oxford.

The authors are grateful for the LIGO Scientific Collaboration review of the paper and this paper is assigned a LIGO DCC number(P2000092). The Columbia Experimental Gravity group is grateful for the generous support Columbia University in the City of New York, for National Science Foundation grant PHY-1708028, and for computational resources provided by the LIGO Laboratory (supported by NSF Grants PHY-0757058 and PHY-0823459). DV is grateful to the Ph.D. grant of the Fulbright foreign student program and IB acknowledges support from University of Florida.

\bibliography{references}

\section{Supplement}
\subsection{Maximum Likelihood Analysis}
\begin{figure}[h!]
    \centering
    \includegraphics[width=0.85\textwidth]{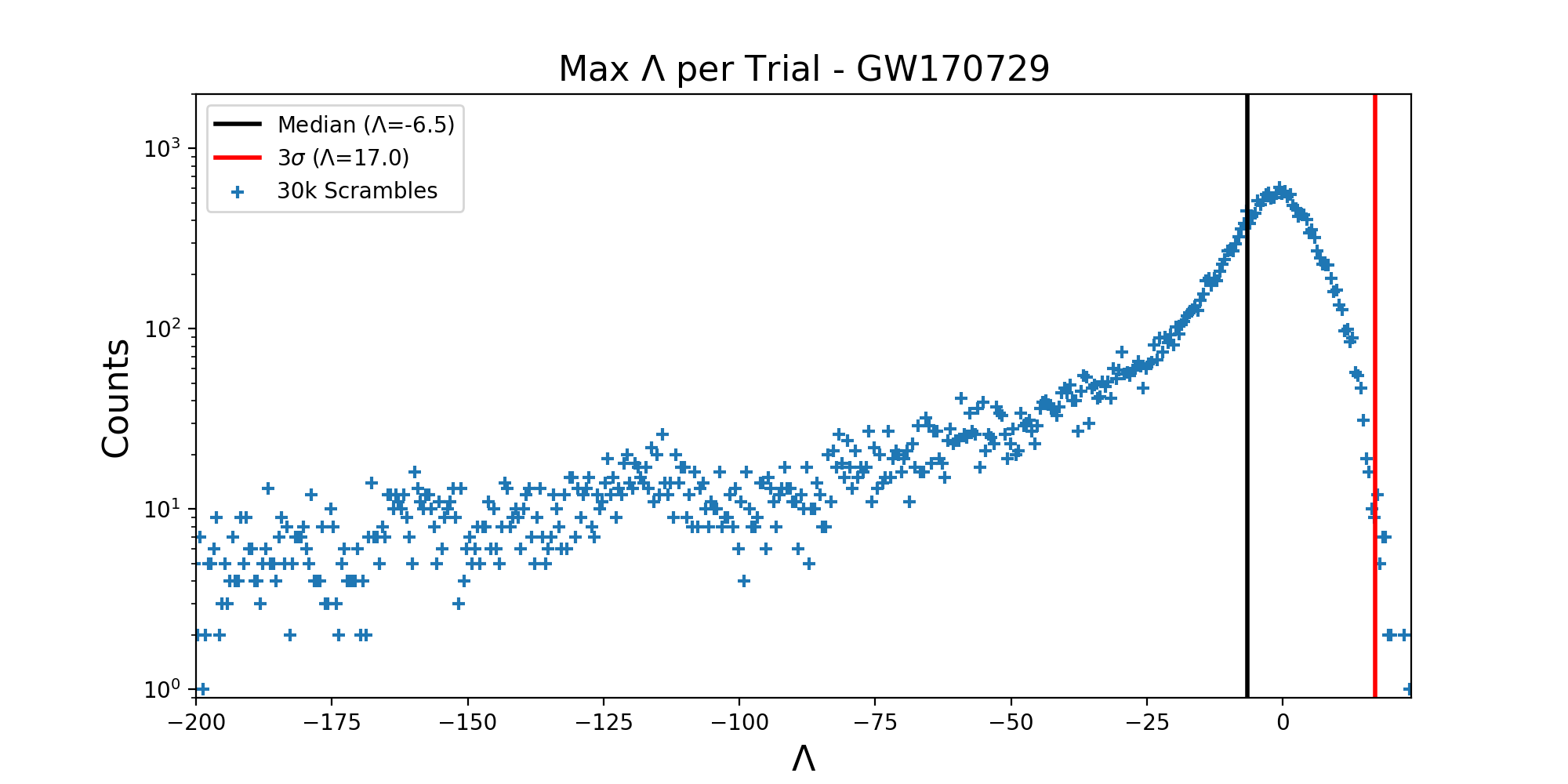}
    \caption{Background test statistic distribution for GW170729. We see a large fraction of $\Lambda$ values that are negative due to most neutrino events on the sky being heavily down-weighted by the spatial prior.}
    \label{fig:bkgTS}
\end{figure}

Figure \ref{fig:bkgTS} shows the background distribution for the unbinned maximum likelihood analysis. Here, 30,000 trials are performed for GW170729. Neutrino arrival times are randomly assigned, which has the effect of assigning a random right ascension value to each neutrino, while preserving the declination and temporal distribution of our data. Since IceCube's sensitivity does not depend on right ascension, we can randomize neutrinos in right ascension to produce different realizations on the sky.

For each trial and random realization of the sky, the likelihood is maximized in each pixel on the sky and the maximum resulting test statistic on the sky is taken as the test statistic for that trial. Shown in Figure \ref{fig:bkgTS} are the maximum TS values on the sky for 30,000 unique trials.

\begin{figure}[htb!]
    \centering
    \includegraphics[width=0.55\textwidth]{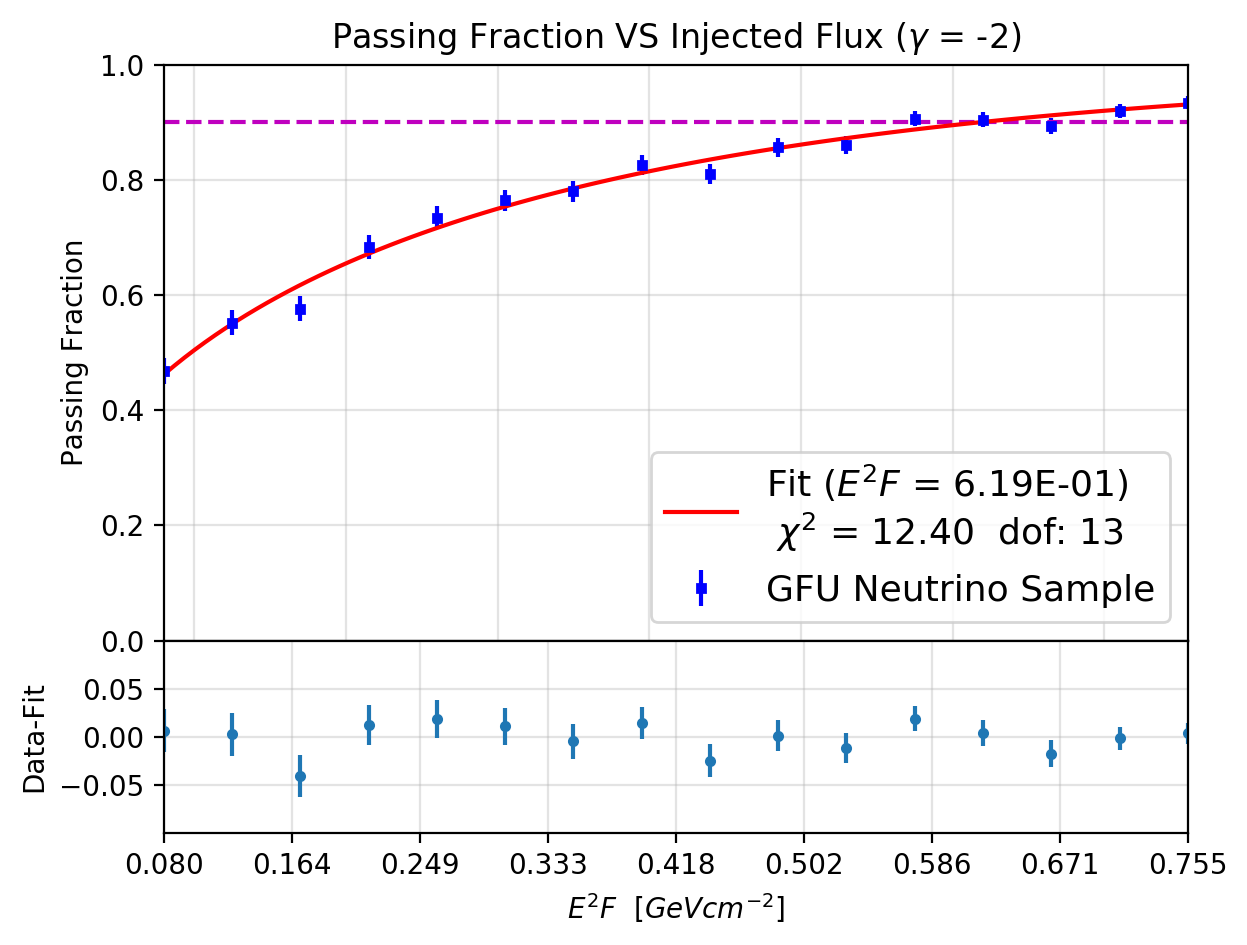}
    \caption{90\% Sensitivity flux for GW170104, computed by injecting an increasing neutrino flux according to an $E^{-2}$ power-law spectrum and calculating the fraction of trials which return an observed $\Lambda$ greater than the median of the background distribution shown in Figure \ref{fig:bkgTS}. We use a $\chi^2$ cumulative distribution function (CDF) to fit the data and compute the time integrated flux which gives a 90\% passing fraction. This is defined as the sensitivity for the particular GW event.}
    \label{fig:sensitivity}
\end{figure}

Figure \ref{fig:sensitivity} shows the method used to compute sensitivity and upper limits in the unbinned maximum likelihood analysis. We perform signal injection trials with neutrinos from Monte Carlo. For each trial we choose a random location on the sky weighted by the GW localization and inject signal. We compute a passing fraction which refers to how many trials yielded a test statistic larger than the background median for a given injected flux. We increase the injected flux and perform trials again. We repeat this procedure for multiple injected flux values and fit a $\chi^2$ CDF to the passing fraction vs injected flux as shown in Figure \ref{fig:sensitivity}. In the case of computing upper limits, the passing fraction refers to how many trials yielded a test statistic larger than the observed test statistic for the GW. We fix the lower bound of the upper limit to the sensitivity to be conservative. 

\subsection{LLAMA}
The fluence upper limits of the LLAMA search in Table \ref{tab:results} are calculated as follows; three sets of neutrino lists are prepared for each GW event. In the first set the list consist of only background neutrinos whose count is drawn from a Poisson distribution with the mean background neutrino count in 1000~s. The neutrinos are chosen from the archival GFU dataset. Neutrinos' right ascensions and detection times are randomized and other properties are kept the same. The detection times of these neutrinos are uniformly chosen around the $\pm500$~s of the GW event time. Right ascensions are uniformly randomized between $[0,2\pi]$. The second set of lists have one signal neutrino in addition to first set of lists. The emission location of the signal neutrinos are randomized according to the GW sky localization. The detection times of the neutrinos are chosen from a symmetric triangular distribution whose mode is the GW event time and extent is $\pm500$~s around the GW event time. The neutrinos are chosen from a Monte Carlo signal simulation list with $E^{-2}$ energy spectrum. The Monte Carlo list consists of neutrinos detected all over the sky. For choosing neutrinos from this list, only the neutrinos in the $\pm1 \degree$ declination band around the emission location are considered. The Monte Carlo list has a real location, a detection location and detection error for the location for each neutrino. Maintaining the difference between the real and detection location, chosen neutrinos' real locations are shifted to emission points which have been chosen on the sky according to the GW's sky localization distribution. The third set of lists are similar to the second set of lists except having two signal neutrinos emitted from the same point in each list. Then the analysis is run on these sets of lists. Finally we find the fluence value for which 90\% of the detected lists of neutrinos will have a higher TS than the TS of the actual detection. The fluence affects the expected number of signal neutrinos. Expected number of signal neutrinos for a fluence are found by using the effective area of the detector and assuming a $E^{-2}$ spectrum. By using the TS values from the three sets of lists we can calculate the fraction of events that would have higher TS than the actual detection's TS for a fluence value. When doing this we assume that the events which have 3 or more signal neutrinos will always have a higher TS than the actual detection's TS since such a detection would be extremely significant. Moreover we put a lower bound to the actual detection's TS which is equal to the median of the TS distribution of only background neutrinos (sensitivity). See \cite{veske2020neutrino} for limits obtained based on maximum likelihood estimators without the sensitivity lower bound for the first 3 GW events.

\begin{figure}[h!]
    \centering
    \includegraphics[width=0.5\textwidth]{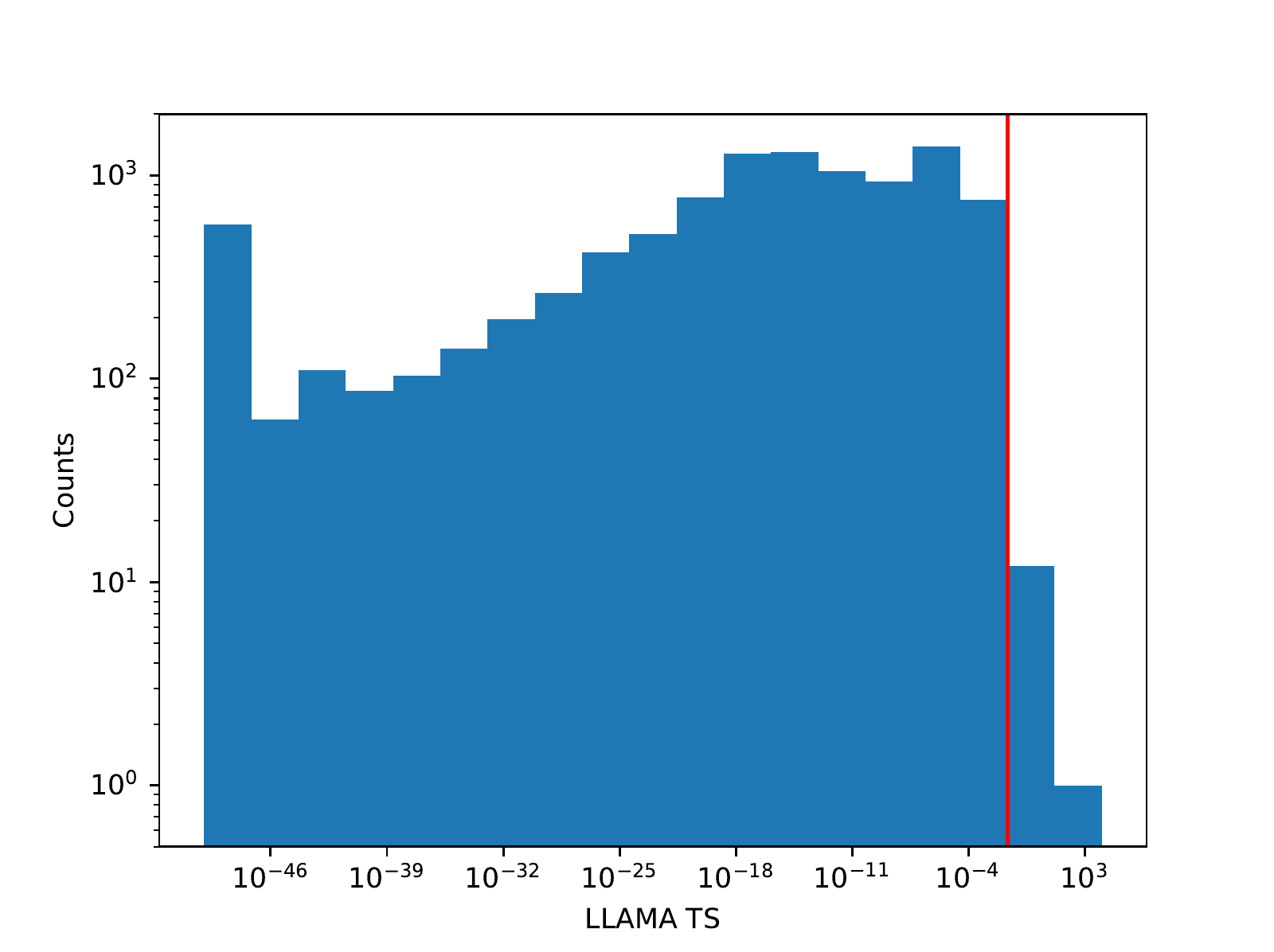}
    \caption{Background TS distribution of the LLAMA search for BBH mergers detected with 2 aLIGO and the AdVirgo detectors at O2 sensitivity. TS $<10^{-50}$ are collected in the $10^{-50}$ bin. The red line shows the one sided 3$\sigma$ threshold.}
    \label{fig:llama_bg_dist}
\end{figure}

Figure \ref{fig:llama_bg_dist} shows the background TS distribution of the LLAMA search used for the analysis of joint GW high-energy neutrino events which have GWs detected by the 3 detector network of 2 aLIGO detectors and the AdVirgo detector. In order to create background pairs of GWs and high-energy neutrino events, first GWs were injected and detected at O1/O2 sensitivity with the 3 GW detectors. The injections were made uniform in comoving volume. The injection volume is as large as the detectors maximum reach. Then each detected GW injection was paired with a set of neutrinos whose count was drawn from a Poisson distribution whose mean was the mean count of the neutrino detections from the GFU stream in 1000 seconds. According to the Poisson draws, that many neutrinos were chosen from the archival GFU dataset. Before the selection the right ascension coordinate of the neutrinos in the dataset were scrambled. Finally those neutrinos were distributed uniformly in the $\pm 500$ s window around the GW injections which created the background joint GW high-energy neutrino event. The search was run on a set of $\sim 10^4$ such events.

\begin{figure}[h!]
\begin{center}
\begin{tabular}{cc}
    \includegraphics[width=0.35\textwidth]{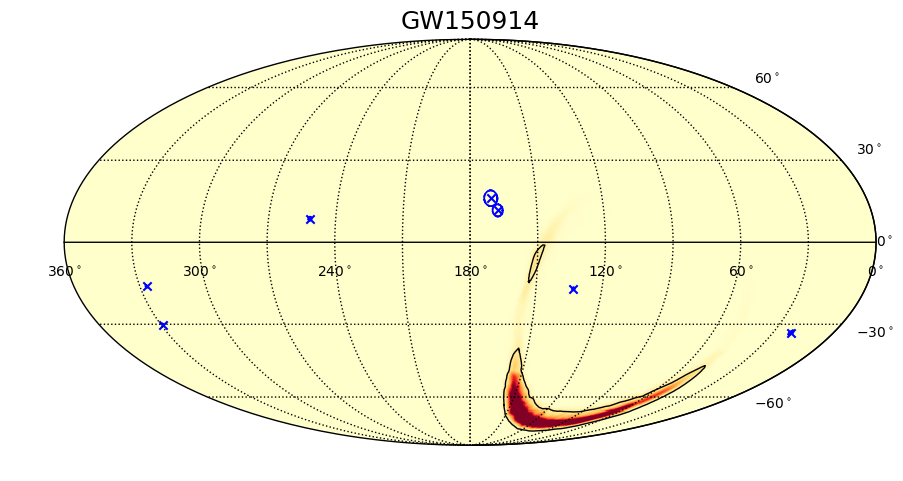} &
    \includegraphics[width=0.35\textwidth]{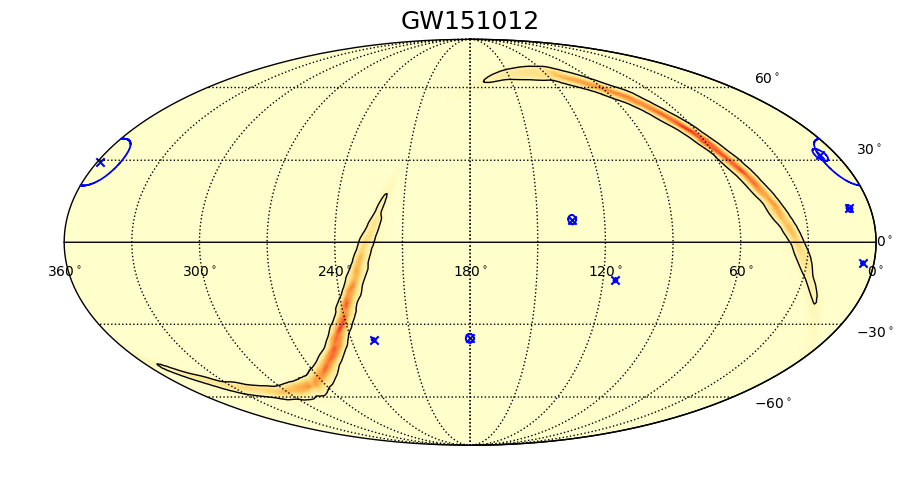} \\
    \includegraphics[width=0.35\textwidth]{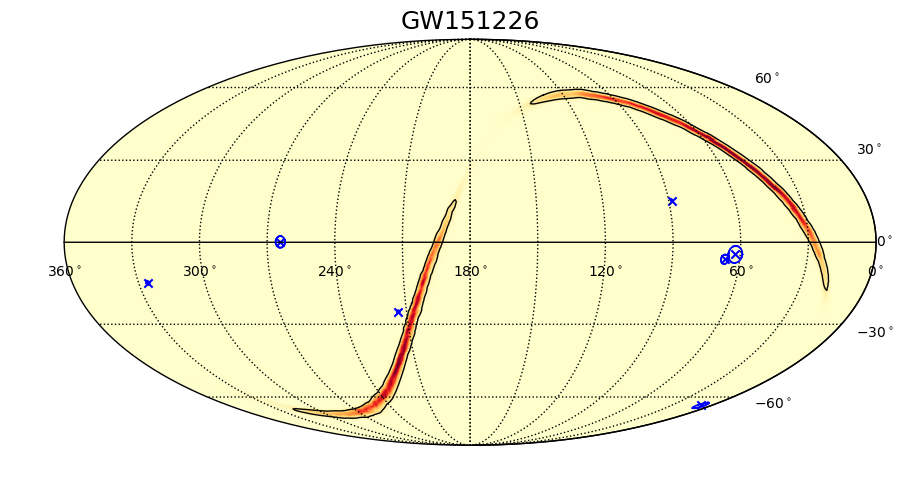} &
    \includegraphics[width=0.35\textwidth]{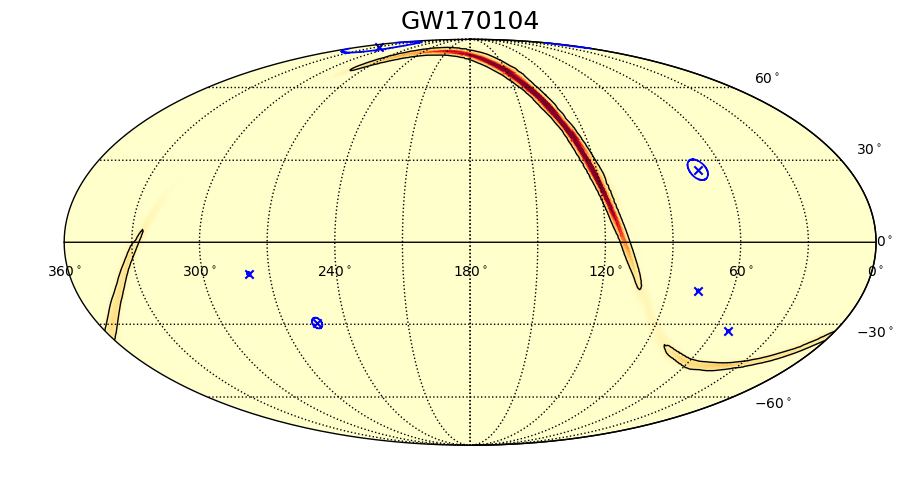} \\
    \includegraphics[width=0.35\textwidth]{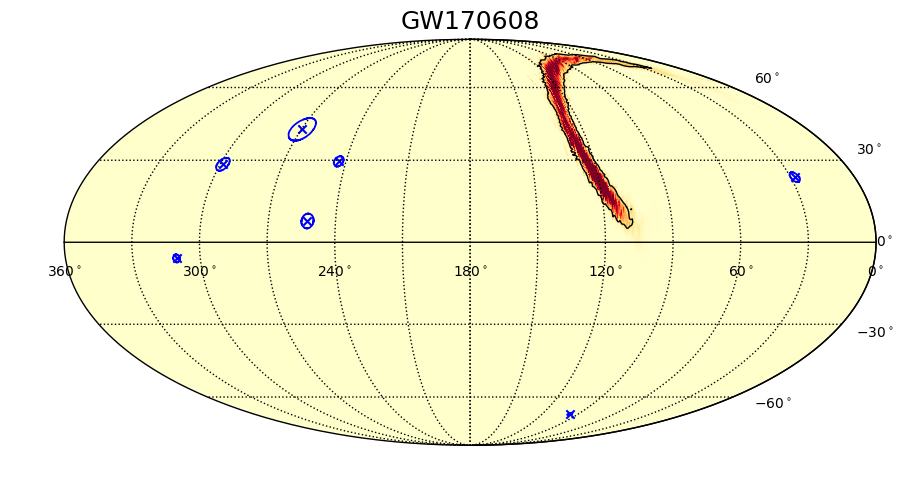} &
    \includegraphics[width=0.35\textwidth]{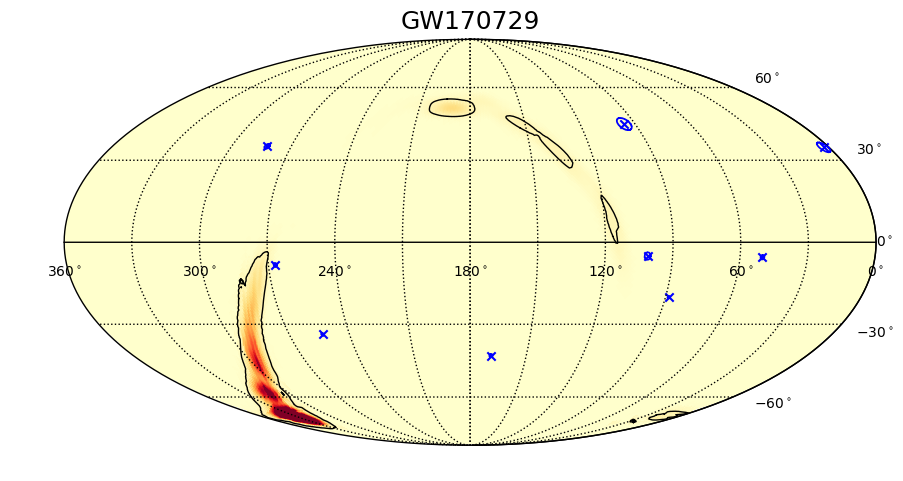} \\
    \includegraphics[width=0.35\textwidth]{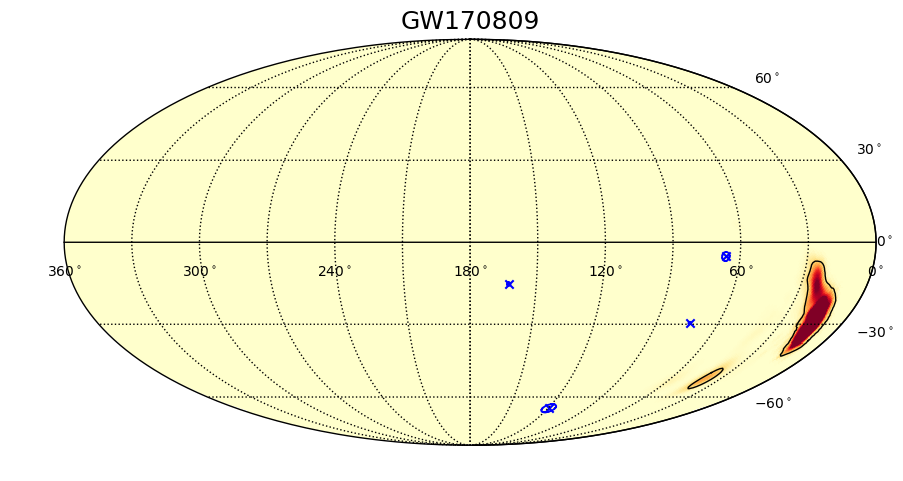} &
    \includegraphics[width=0.35\textwidth]{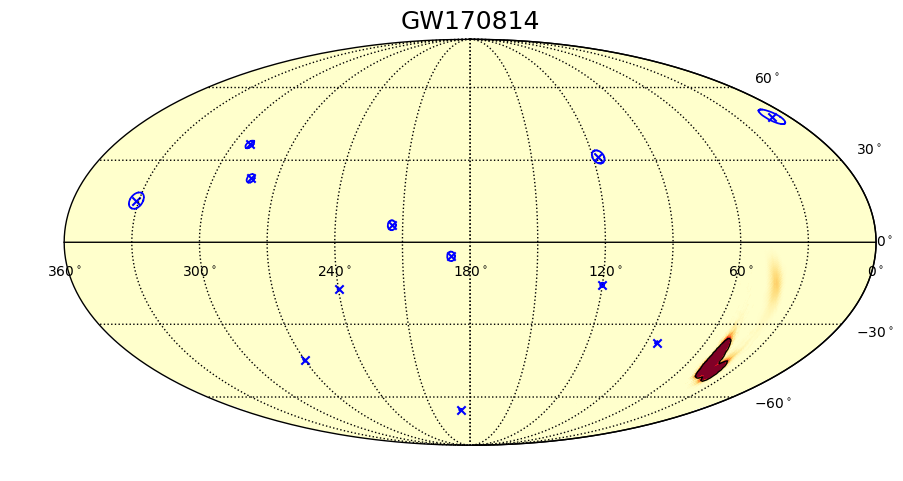} \\
    \includegraphics[width=0.35\textwidth]{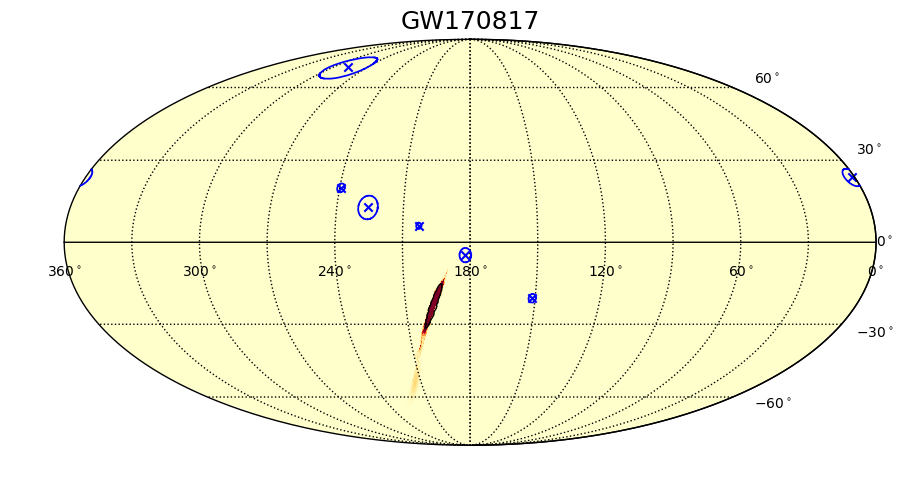} &
    \includegraphics[width=0.35\linewidth]{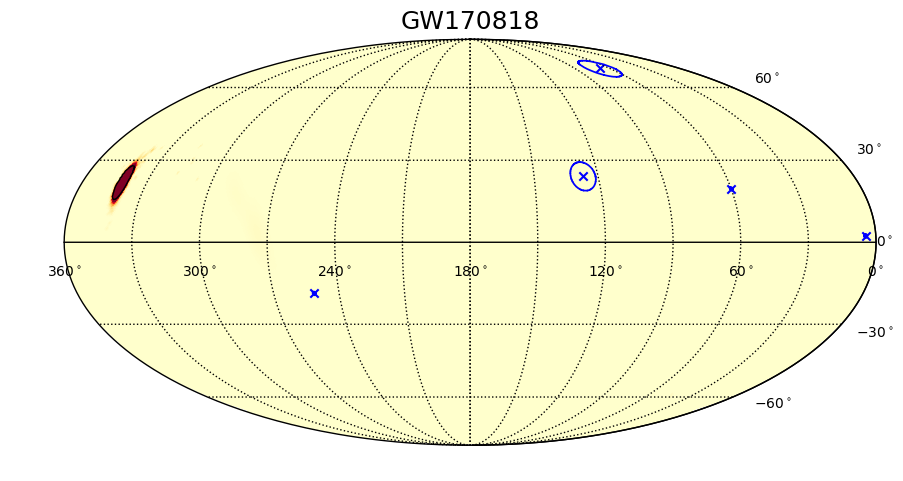} \\
    \multicolumn{2}{c}{\includegraphics[width=0.40\linewidth]{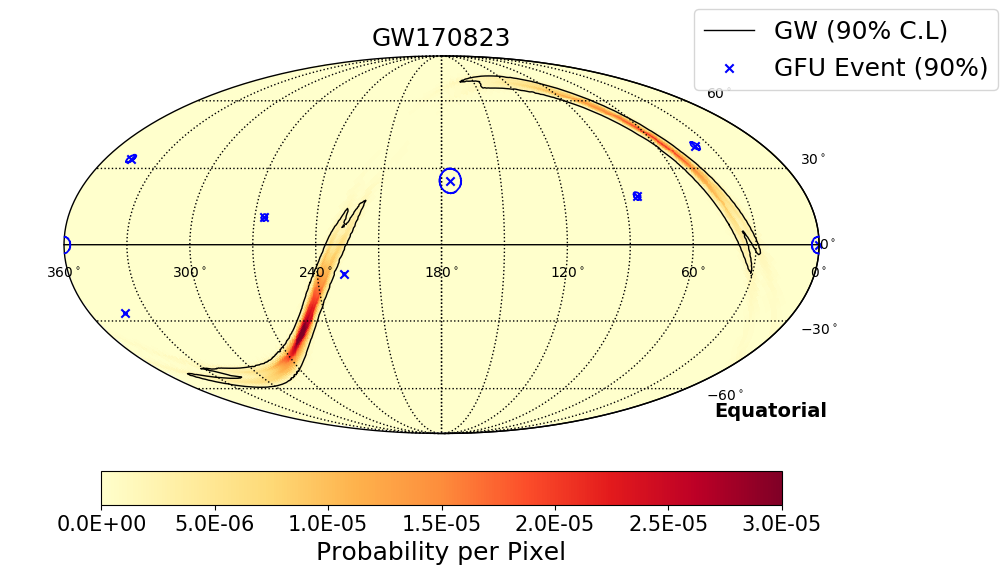}} 
\end{tabular}
\end{center}
\caption{Skymaps for the 11 detected GW events overlayed with neutrinos within 1000 seconds of the GW trigger time.}
\label{fig:skymap_gallery}
\end{figure}

\end{document}